# Dynamic Voltage Stiffness Control Technique for a Virtual Oscillator based Grid-forming Controller

Ritwik Ghosh, Narsa Reddy Tummuru, *Senior Member, IEEE*, and Bharat Singh Rajpuohit, *Senior Member, IEEE*

*Abstract*— Virtual oscillator control is the latest control technique for grid-forming inverters. Virtual Oscillator based Controllers (VOCs) provide all the steady-state droop functionalities of conventional droop controllers and, in addition, the time-domain synchronization with a connected electrical network. However, existing literature does not consider the aspect of dynamic control over the voltage stiffness of a VOC. Voltage stiffness is a vital parameter for a grid-forming inverter. If the voltage stiffness is too high, the inverter picks up all the reactive power demand of the PCC. In contrast, if the stiffness is too low, the inverter does not participate in voltage regulation at all. Limiting the reactive power output during a higher voltage sag, especially when connected to a weak grid, is challenging for a VOC. Entering into the current control mode is the existing solution, but it severely affects the effective synchronization between the VOC and the voltages of the PCC. As a result, the grid-forming mode of operation becomes inefficient. This article has introduced a Virtual Impedance (VIm) based dynamic voltage stiffness control technique for VOCs. The systematic design procedure for the proposed voltage stiffness controller is presented. In addition, a rigorous approach for stability analysis is presented.

## I. INTRODUCTION

Non-linear limit cycle oscillator (Virtual Oscillator) based control is the latest control strategy for the grid-forming inverters [1]–[3]. Virtual Oscillator (VO) control meets all the steady-state functionalities of conventional droop control and virtual synchronous generator control but with improved dynamic performance [4], [5]. Among different models of oscillators, the Andronov-Hopf oscillator and Dispatchable Virtual Oscillator (d-VOC) based controllers are proved to be the best fits for grid-forming controllers [6], [7]. The elemental form of an Andronov-Hopf oscillator and a d-VOC is similar [6]. A grid-forming controller should meet the system-level requirements such as fault ride-through capability, MPPT capability, and capability to perform under unbalanced grid conditions [8]. The recent research works [9]–[11] have provided a very general VO-based control architecture where all the above-mentioned system-level functionalities are included.

However, the existing literature on VOCs does not consider the dynamic control over the voltage stiffness. Commonly a VOC is designed for less than a 10% tolerable voltage sag range [6]. The reason behind a shorter low voltage bound is that the control over the reactive and active power droop characteristics of VOCs are not decoupled. The reactive power droop cannot be changed independently without changing the active power droop. It is important to mention that voltage stiffness is an important parameter for any grid-forming inverter, especially when connected to a weak grid [8]. During the addition of a large load in an electrical network, the grid-forming inverter, which is electrically nearest, is affected first [12]. If the connected grid is weak, the voltage drop becomes large. Hence, the reactive power demand on the nearest grid-forming inverter becomes significantly high. At this point, if the voltage stiffness of the grid-forming inverter is too low, the inverter does not participate in voltage regulation at all [8]. As the grid is weak, the voltage amplitude drops further. On the other hand, if the voltage stiffness of the inverter is too high, the inverter picks up all the reactive power demand on itself [8], and the output current may tend to exceed the maximum rating. When the output current tends to the limit, the existing VOCs have two options. The first option is to be disconnected from the electrical network. However, the new grid codes prevent grid-forming converters from being disconnected easily from the network in such a condition as mentioned above [13]. The second option is to enter into current control mode by activating the fault ride-through controller [9], [14]. However, the effective synchronization of a grid-forming controller with the connected network gets severely affected in the current control mode [8]. The exact reason behind the loss of synchronization of a VOC in the current control mode is investigated using a simulation study in Section IV. It is not a very good choice to enter into the current control mode during every voltage sag caused by a load transient at the PCC. Also, a discrete and higher jump in the value of the virtual resistor, as presented in [14] reduces the effective X/R ratio in the resultant source impedance of the grid-forming inverter. A lower X/R ratio cannot ensure superior decoupling between the active and reactive power output of the inverter [12].

Virtual impedance (VIm) based control technique has been a popular choice for voltage-source and current-source converters in recent years [15]. VIm-based fault ride-through technique is presented in [16], [17] for grid-forming converters. The converter enters into the grid-following mode when the current limiting controller is activated. However, a fault ride-through technique is not suitable for dealing with typical voltage sags (10-20%) frequently occurring in power systems [18]. A VIm-based control technique is used for improving the reactive power sharing accuracy among parallelly connected grid-forming inverters in AC islanded microgrids in [19], [20]. However, the requirement of communication links makes the control strategies proposed in [19], [20] less reliable. VIm-based decentralized control technique for accurate reactive power sharing among parallelly connected grid-forming inverters in AC islanded microgrids is presented in [12], [21]. VIm-based controllers are used in [22], [23] to mitigate voltage unbalance in the islanded microgrid. The proposed method presented in [22] requires a secondary centralized controller, which reduces the reliability of the system. The literature [23] considers a maximum voltage unbalance of 8.2%, whereas a voltage unbalances of 2-15% for a duration of 3 cycles to 1 minute is very typical in electrical power systems [18].

This article introduces a VIm-based dynamic voltage stiffness control method for VO-based grid forming controllers. The discrete research contributions of this article are summarized as follows.

1. This article has presented a simulation study and analysis to point out the reason behind the loss of effective synchronization of a VOC in the current control mode.

2. The proposed dynamic voltage stiffness controller provides an extra degree of freedom to a VO-based grid-forming controller to modify the reactive power droop characteristic without changing the active power droop characteristic.

3. The proposed controller dynamically vary the virtual impedance to ensure a lower voltage stiffness under normal condition and a higher voltage stiffness under voltage sags while maintaining a higher X/R ratio throughout the operating range.

4. The proposed controller has decoupled control over individual phases. It helps the VOC to withstand significantly larger unbalance voltage sag (more than 15%).

5. Using the proposed controller, an existing VOC can maintain effective synchronization with a connected electrical network during balanced and unbalanced sags and remain in the grid-forming mode under significantly higher voltage sags (more than 20%).

6. The proposed controller can be integrated into an existing VO-based system-level grid-forming controller without the requirement of any change in the existing controller or any extra sensor, as shown in Fig. 1.

7. It is challenging to design the parameters of the proposed adaptive VIm-based voltage stiffness controller. A large and fixed value of virtual impedance can increase the range of voltage sag that a grid-forming inverter can withstand. However, it decreases the participation towards voltage regulation from the inverter. This is why instead of putting a fixed virtual impedance, the proposed technique adopts variable virtual impedance. This article has presented a detailed design procedure that can achieve a suitable trade-off between the increase in tolerable voltage sag limit and a decrease in the reactive power output.

8. This article has introduced a rigorous approach of small signal stability analysis for VO-based grid-forming controller where the detailed model of the power system, voltage dynamic of the VOC, nested voltage and current loops, and the proposed controller are taken into account.

9. A brief overview is presented in the Conclusion on how the present research work will be extended in the future to provide newer functionalities such as improvement in reactive power sharing among parallelly connected converters.

The rest of the paper is organized as follows. Section II introduces the proposed adaptive VIm-based dynamic voltage stiffness controller. Section II also presents the detailed design procedure for the key parameters of the proposed controller. In Section III, a small-signal analysis is presented to find the range of the key parameters of the proposed controller. Simulation and experimental studies are presented in Section IV and Section V to validate the functionalities offered by the proposed VIm method when integrated into a VO-based grid forming controller. Finally, the paper is concluded in Section VI with a brief overview of how the present research work will be extended in the future.

II. THE PROPOSED VOLTAGE STIFFNESS CONTROLLER

The proposed dynamic voltage stiffness controller is integrated into the existing VO-based system level grid-forming controller [11] as shown in Fig. 1. The proposed controller is indicated using red lines and words. The existing grid-forming controller is presented in detail in [11].

The proposed controller has the two following objectives

(i) The voltage stiffness of a grid-forming inverter should be lower when the reactive power demand on the inverter is low. Conversely, the voltage stiffness should be dynamically increased with the increase in the reactive power demand on the inverter.

(ii) A high X/R ratio in the total resultant impedance should be maintained throughout the operating range

The two mentioned objectives are achieved by dynamically varying the virtual impedance as

$$Z_{VIm\_abc} = (R_{V0} + mI_{inv\_abc\_q}) + j(X_{V0} + nI_{inv\_abc\_q}) \quad (1)$$

where $R_{V0}$ and $X_{V0}$ are the initial value of the virtual resistance and reactance, respectively. The parameters, $m$, and $n$ are the adaptive virtual resistance and inductive reactance gain in Ω/A. The parameter, $I_{inv\_abc\_q}$ represents the quadrature axis phase currents of the inverter.

A. *Selecting the value of $R_{V0}$, $X_{V0}$, $m$, and $n$*

The initial virtual resistance and reactance, $R_{V0}$ and $X_{V0}$ are more important for improving reactive power sharing accuracy among the inverters in islanded microgrids [12]. Since this article focused on the voltage stiffness control in dispatchable mode, the initial resistance and reactance are taken as zero

$$R_{V0} = 0; \; X_{V0} = 0 \quad (2)$$

To maintain the same X/R ratio throughout the operating range, the values of $m$ and $n$ are taken as the same

$$m = n = k \quad (3)$$

The active and the reactive power droop characteristic of a symmetrical component based VOC (S-VOC) is represented by [24]

$$\omega_z = \omega_n - \frac{k_v k_i}{CV_z^2}(P_z - P_z^*) \quad (4)$$

$$0 = \frac{\xi}{k_v^2}V_z(2V_{zn}^2 - 2V_z^2) - \frac{k_v k_i}{CV_z}(Q_z - Q_z^*) \quad (5)$$

where, $z \in a, b, c$.

The reactive power droop characteristic can be modified only by changing the current scaling factor, $k_i$. However, changing the value of $k_i$ simultaneously affect the active power droop characteristic.

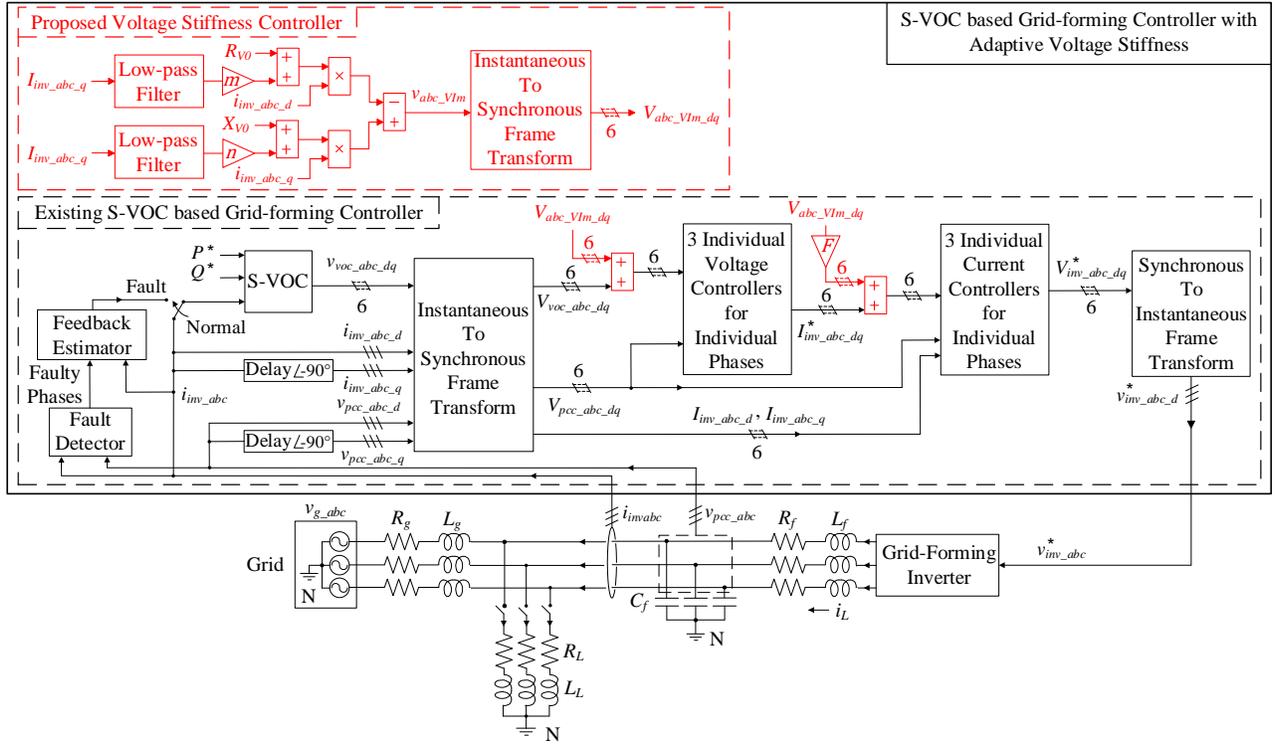

Fig. 1. The schematic diagram of the proposed S-VOC based grid forming controller

As presented in (1), the virtual impedance is a function of quadrature axis phase currents, $I_{inv\_abc\_q}$ only. Hence, with a proper X/R ratio the virtual impedance can modify the reactive power droop without any significant effect on the active power droop. The resultant reactive power droop is derived as

$$Q_{Z\_VIm} = \frac{Q_Z}{\sqrt{1 + \left(\frac{kQ_Z}{ZV_z}\right)^2}} \quad (6)$$

where, $Q_Z$ and $Q_{Z\_VIm}$ are the reactive power output without and with the virtual impedance, respectively. The parameter $Z$ is the total source impedance which is the result of the addition of the filter and grid impedance. As shown in Fig. 2, the reactive power droop characteristic can be modified using the virtual impedance.

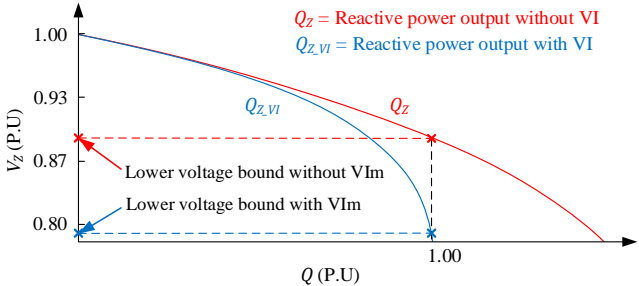

Fig. 2. The reactive power droop characteristic without and with virtual impedance

With a positive value of virtual impedance gain, $k$ the droop characteristic shifts towards the left. The mentioned modification increases the tolerable voltage sag limit. The reactive power output remains nearly the same near the nominal voltage. Conversely, the reactive power output decreases with the increase in voltage sag. The value of $k$ is selected to achieve a proper trade-off between the increase in tolerable voltage sag limit and the decrease in the reactive power output.

### III. SMALL SIGNAL STABILITY ANALYSIS

A stability analysis is required to derive the limiting value of the parameter, $k$. This article has introduced a rigorous approach for small signal stability analysis of a VO-based grid-forming controller. The detailed model of the system and the controller is derived and considered for the stability analysis. The output phase voltage (i.e. $v^*_{voc\_a\_d}$, $v^*_{voc\_b\_d}$, and $v^*_{voc\_c\_d}$) of the S-VOC is taken as the reference for the same individual phase for the instantaneous to synchronous reference frame transformation, as shown in Fig. 3. The instantaneous and synchronous frame-based parameters are denoted by small and capital letters, respectively. The detailed derivation of the reference frame transformation is presented in [11].

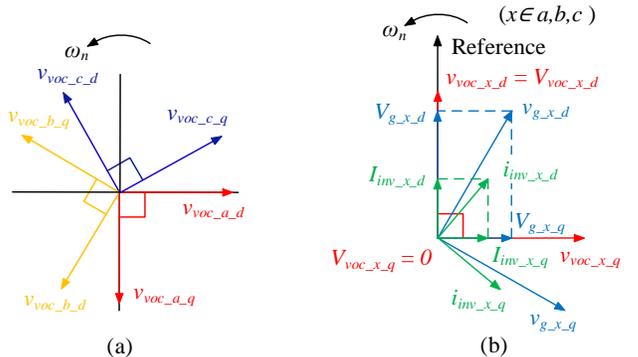

Fig. 3. (a) Output voltage vectors of the S-VOC (b) Instantaneous to synchronous reference frame transformation

The model of the system and the controller is derived in the s-domain as follows.

The model of the connected electrical network is derived in terms of grid impedance and grid voltages as

$$I_{inv\_d} = \frac{\omega L_g}{(R_g + sL_g)} I_{inv\_q} + \frac{1}{(R_g + sL_g)} (V_{pcc\_d} - V_{g\_d}) \quad (7)$$

$$I_{inv\_q} = \frac{-\omega L_g}{(R_g + sL_g)} I_{inv\_d} + \frac{1}{(R_g + sL_g)} (V_{pcc\_q} - V_{g\_q}) \quad (8)$$

The model of the L-C filter of the inverter is derived as

$$I_{L\_d} = \frac{\omega L_f}{(R_f + sL_f)} I_{L\_q} + \frac{1}{(R_f + sL_f)} (V_{C\_d} - V_{pcc\_d}) \quad (9)$$

$$I_{L\_q} = -\frac{\omega L_f}{(R_f + sL_f)} I_{L\_d} + \frac{1}{(R_f + sL_f)} (V_{C\_q} - V_{pcc\_q}) \quad (10)$$

$$V_{pcc\_d} = \frac{\omega}{s} V_{pcc\_q} + \frac{1}{sC_f} I_{C\_d} \quad (11)$$

$$V_{pcc\_q} = -\frac{\omega}{s} V_{pcc\_d} + \frac{1}{sC_f} I_{C\_q} \quad (12)$$

$$I_{C\_d} = I_{L\_d} - I_{inv\_d} \quad (13)$$

$$I_{C\_q} = I_{L\_q} - I_{inv\_q} \quad (14)$$

The model of the virtual impedance is derived as

$$I_{F\_inv\_d} = \frac{\omega_f}{s^2 + \omega_f s + \omega_f^2} I_{inv\_d} \quad (15)$$

$$I_{F\_inv\_q} = \frac{\omega_f}{s^2 + \omega_f s + \omega_f^2} I_{inv\_q} \quad (16)$$

$$R_{VIm} = R_{V0} + mI_{F\_inv\_q} \quad (17)$$

$$X_{VIm} = X_{V0} + nI_{F\_inv\_q} \quad (18)$$

The feed-forward term is included in the model as

$$V_{VIm\_d} = -R_{VIm} I_{F\_inv\_d} - X_{VIm} I_{F\_inv\_q} \quad (19)$$

$$V_{VIm\_q} = -R_{VIm} I_{F\_inv\_q} + X_{VIm} I_{F\_inv\_d} \quad (20)$$

The model of the voltage controller is derived as

$$I^*_{L\_d} = \left(K_{pv} + \frac{K_{iv}}{s}\right)(V_{voc\_d} + V_{VIm\_d} - V_{PCC\_d}) - FV_{VIm\_d} \quad (21)$$

$$I^*_{L\_q} = \left(K_{pv} + \frac{K_{iv}}{s}\right)(V_{voc\_q} + V_{VIm\_q} - V_{pcc\_q}) - FV_{VIm\_q} \quad (22)$$

The model of the current controller is derived as

$$V_{C\_d} = \left(K_{pi} + \frac{K_{ii}}{s}\right)(I^*_{L\_d} - I_{inv\_d} - \omega C_f V_{pcc\_q}) - \omega L_f V_{pcc\_q} \quad (23)$$

$$V_{C\_q} = \left(K_{pi} + \frac{K_{ii}}{s}\right)(I^*_{L\_q} - I_{inv_q} + \omega C_f V_{pcc\_d}) + \omega L_f V_{pcc\_d} \quad (24)$$

The complete small-signal model of the system and the controller is shown in Fig. 4. The S-VOC is intended to maintain effective synchronization. When the S-VOC is synchronized, it can be approximated that

$$V_{voc\_q} \approx V_{voc\_q} \approx V_{C\_q} \approx 0 \quad (25)$$

The main interest is to find the relation between $I_{L\_q}$ and $V_{pcc\_d}$ i.e. $(I_{L\_q}/V_{pcc\_d})$ in terms of virtual impedance gain, $k$. The expression of $(I_{L\_q}/V_{pcc\_d})$ is linearized around the point $(Q_{z\_max}, V_{z\_min})$. The other known parameters are derived as follows.

The range of the proportional and integral gains of the current and the voltage controller is given as follows [10]

$$K_{pi} = L_f \omega_i; \; K_{ii} = R_f \omega_i \quad (26)$$

$$K_{pv} = C_f \omega_v; \; K_{iv} = \frac{2K_{pv}\omega_v^2}{\omega_i} \quad (27)$$

where $\omega_i$, $\omega_v$ are the current and voltage loop bandwidth. The value of $\omega_i$, $\omega_v$ are taken $2\pi 1500$ rad/s and $2\pi 400$ rad/s [10]. The values of the known parameters are given in Table I. The system and control parameters are taken directly from standard references, [6] and [10].

The eigenvalue loci for the virtual impedance gain, $k$ are presented in Fig. 5. For the given system, the limiting value of $k$ is found to be 0.22 Ω/A.

Table I: Specifications of the system and controller parameters for the simulation study

| Symbol | Description | Value |
|---|---|---|
| | Power System | |
| $V_{ng}$ | Nominal phase voltage (Grid) | 80 V |
| $\omega_{ng}$ | Nominal frequency | $2\pi 60$ rad/s |
| $L_g$ | Grid side inductor | 4 mH/Phase |
| $R_g$ | Grid side resistor | 0.35 Ω/Phase |
| $L_f$ | Filter inductor | 2 mH/Phase |
| $R_f$ | Filter resistor | 0.15 Ω/Phase |
| $C_f$ | Filter capacitor | 20 μF/Phase |
| $R_L, L_L$ | R-L Load at the PCC | (2.5 Ω+10 mH)/Phase |
| | Controller | |
| $k_i$ | Current scaling of S-VOC | 0.1 V/A |
| $\xi$ | Speed constant of S-VOC | 15 (Vs)$^{-2}$ |
| $\omega_v$ | $V_{dq}$ control bandwidth | $2\pi 400$ rad/s |
| $\omega_i$ | $I_{dq}$ control bandwidth | $2\pi 1500$ rad/s |
| $k_{pv}$ | Proportional gain: voltage controller | 0.5 (Ω)$^{-1}$ |
| $k_{iv}$ | Integral gain: voltage controller | 67 (Ωs)$^{-1}$ |
| $k_{pi}$ | Proportional gain: current controller | 5 Ω |
| $k_{ii}$ | Integral gain: current controller | 250 Ωs$^{-1}$ |

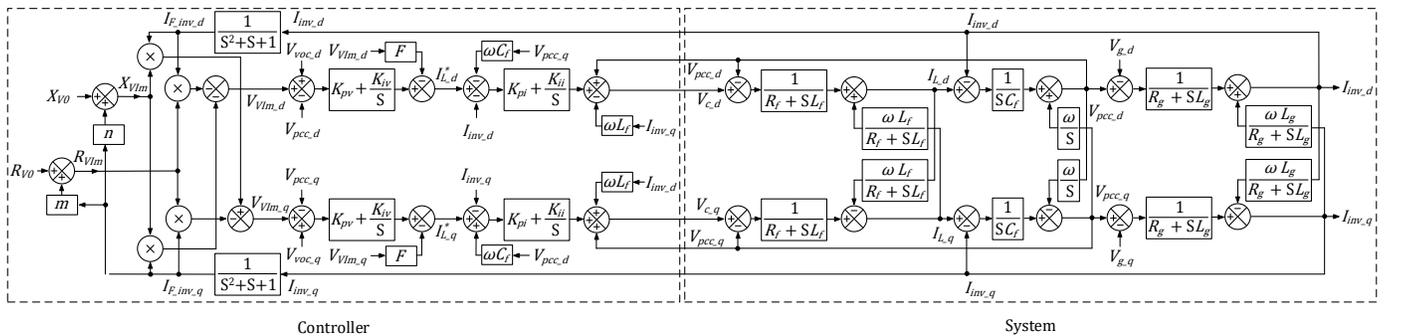

Fig. 4. Small signal model of the system and the controller

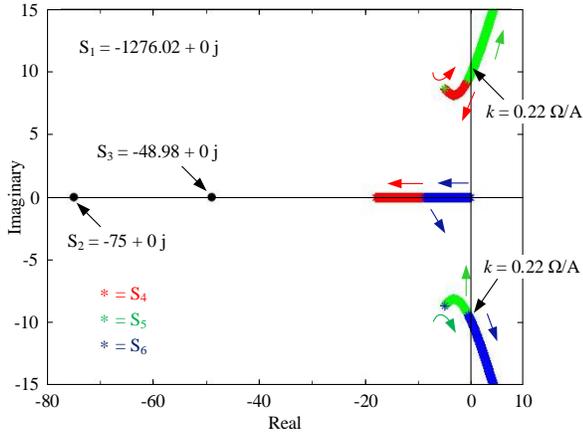

Fig. 5. Eigenvalue loci for the impedance gain, $k$

## IV. THEORETICAL VALIDATION USING SIMULATION STUDY

Two following functionalities of the proposed VIm-based dynamic voltage stiffness controller are validated using simulation studies in this section.

1. To stay operating in the grid forming mode under large voltage sag in a weak grid situation by controlling the voltage stiffness.
2. To achieve the voltage stiffness control without any significant effect on the active power droop characteristic of the VOC.

The schematic diagram of the system considered for the simulation studies is presented in Fig. 1. The parameters of the system are given in Table I. The grid impedance is taken higher to mimic a weak grid situation. At t = 5 s a large load, (2.5 Ω + 2 mH)/phase is added to the PCC. Without any reactive power support from the grid-forming inverter, the voltage amplitude of the PCC drops by 24.81%. Next, the grid-forming inverter without and with the proposed voltage stiffness controller is connected to the PCC. The active power reference, $P_Z^*$ per phase is set to 300W.

The performance of the VO-based grid-forming controller without the proposed voltage stiffness controller is shown in Fig. 6. The output current of the inverter tends to reach the over-current limit without the voltage stiffness controller. As a result, the anti-windup current limiter is activated and restricts the output current under the over-current limit. However, a large oscillation occurs in active and reactive power output. The reason for the power oscillation is as follows. Unlike a droop controller or virtual synchronous machine controller, a VOC uses the instantaneous inverter currents for asymptotic synchronization with a connected network. The output current of the inverter should be the result of the interaction between the phase voltage of the VOC and the PCC for effective synchronization. Once the anti-windup function is activated, the mentioned condition for effective synchronization is no longer satisfied. The anti-windup controller entirely controls the output current of the inverter.

Next, the proposed adaptive voltage stiffness controller is used with a virtual impedance gain, $k$ of 0.07 Ω/A. The virtual impedance modifies the quadrature axis current output of the inverter, as shown in Fig. 7. The output current no longer exceeds the over-current limit. Therefore, the inverter can stay operating in grid-forming mode, as shown in Fig. 8. and keep supporting the PCC with reactive power. The voltage profile of the PCC is improved (from 24.8 % sag to 17.19% sag) by the support of the inverter.

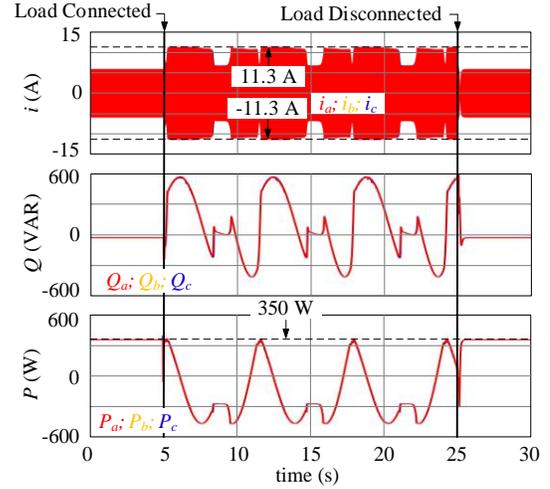

Fig. 6. Performance of a VO-based grid forming controller in the presence of large voltage sag without the proposed voltage stiffness controller

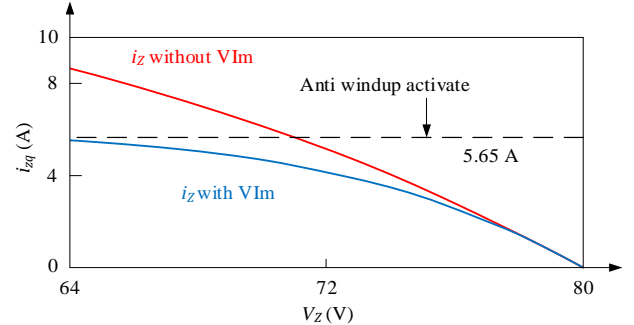

Fig. 7. Quadrature axis current output of the inverter without and with the virtual impedance gain

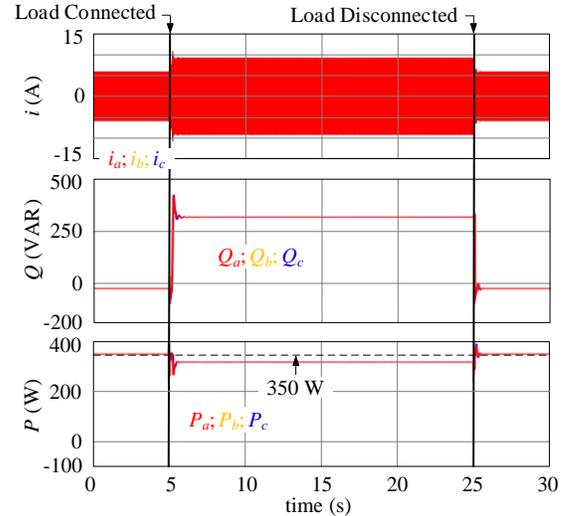

Fig. 8. Performance of a VO-based grid forming controller in the presence of large voltage sag with the proposed voltage stiffness controller

Finally, the active power droop response of the inverter without and with the added virtual impedance gain is obtained. The frequency of the grid is changed from 60 Hz to 59 Hz in five discrete steps, and the active power output of the inverter is plotted in Fig. 9. The voltage of the PCC is kept at the nominal value. The Active power droop characteristic of the inverter without and with the VI are nearly the same and follow the desired active power droop characteristic closely.

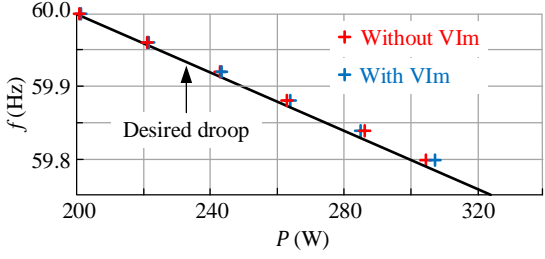

Fig. 9. Active power droop response of the VOC without and with the proposed voltage stiffness controller

## V. EXPERIMENTAL RESULTS AND DISCUSSIONS

The following experiments describe how the proposed VIm-based dynamic voltage stiffness control technique can modify the inherent reactive power droop of an S-VOC without affecting the active power droop. The schematic diagram and the picture of the experimental setup are shown in Fig. 10 and Fig. 11, respectively. A grid emulator, as depicted in Fig. 10, is used. A constant amplitude and frequency inverter, $Inv_2$, behind an inductance, $L_{g1}$, acts as the grid. The frequency and voltage amplitudes of $Inv_2$ are controllable. The grid emulator acts as an ideal sink or source throughout the operating range, i.e., 25 A continuous rms current. The three-phase rectifier, $Rec_2$, which is connected to the utility grid, energizes the dc-link ($V_{dcp2}$, $V_{dcn2}$) of the inverter, $Inv_2$. The specification of the experimental setup is presented in Table II. The OPAL-RT real-time controller (OP4510) is used to deploy the control strategy.

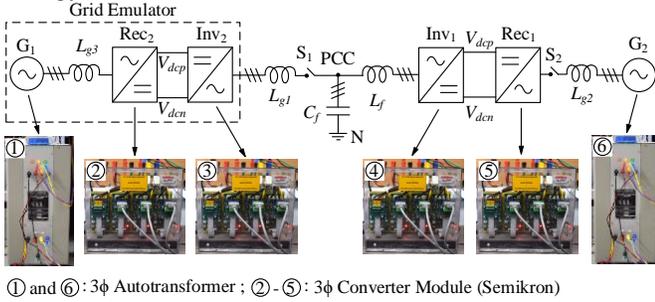

① and ⑥ : 3φ Autotransformer ; ② - ⑤ : 3φ Converter Module (Semikron)

Fig. 10. The schematic diagram of the experimental setup

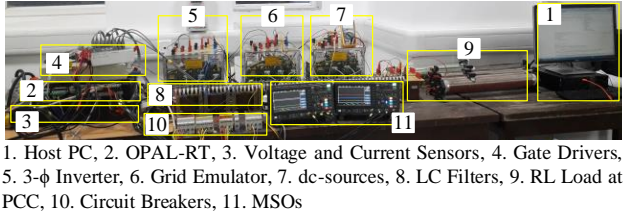

1. Host PC, 2. OPAL-RT, 3. Voltage and Current Sensors, 4. Gate Drivers, 5. 3-φ Inverter, 6. Grid Emulator, 7. dc-sources, 8. LC Filters, 9. RL Load at PCC, 10. Circuit Breakers, 11. MSOs

Fig. 11. The picture of the experimental setup

Table II: Specifications of the experimental setup

| Symbol | Description | Value |
|---|---|---|
| $V_{ng}$ | Nominal voltage: grid emulator | 100 V |
| $\omega_{ng}$ | Nominal frequency: grid emulator | $2\pi 50$ rad/s |
| $L_f$ | Filter inductor: sources | 2 mH/Phase |
| $C_f$ | Filter capacitor: sources | 20 µF/Phase |
| $V_{dcp}, V_{dcn}$ | dc-link voltages (split Capacitor) | 200 V |
| $T_s$ | Sampling time of the controller | 50 µs |

### A. Dispatchable operation

At first, the normal dispatchable operation is conducted. The aim is to observe if there is any adverse effect on the normal operation due to the virtual impedance. The virtual impedance gain, $k$ is set to 0.1 Ω/A. The three-phase active power reference, $P^*$, is initially set to 1500 W. The phase currents and the active power outputs of the inverter are shown in Fig. 12. Initially, each phase injects 500 W of active power into the grid emulator. Then the reference is increased to 2400 W. It is observed that the controller successfully tracks the active power reference.

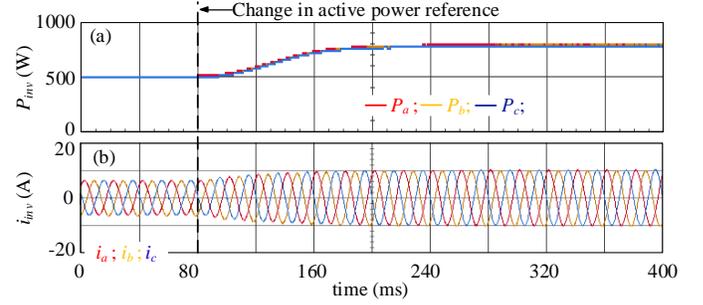

Fig. 12. Normal dispatchable operation of the grid-forming inverter: (a) active power outputs of the individual phases (b) phase currents of the inverter

### B. Effect of virtual impedance on reactive power output

The reactive power output of the inverter is observed under two different voltage sags (5% and 25%) with two different values of virtual impedance gain (0.05 Ω/A and 0.15 Ω/A). The reactive power outputs of the inverter in the mentioned conditions are shown in Fig. 13 and Fig. 14.

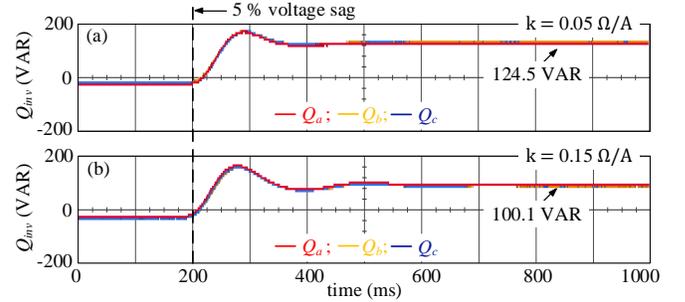

Fig. 13. Reactive power support by the inverter during 5% voltage sag: (a) virtual impedance gain, $k = 0.05$ Ω/A, (b) virtual impedance gain, $k = 0.15$ Ω/A

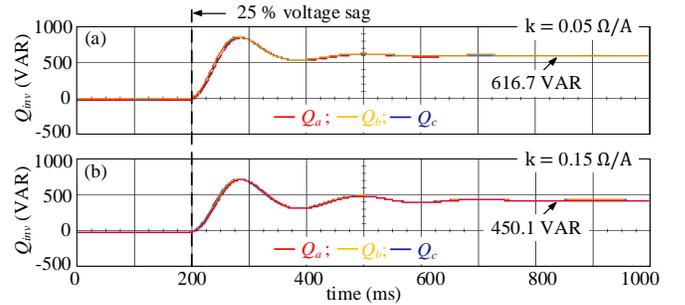

Fig. 14. Reactive power support by the inverter during 25% voltage sag: (a) virtual impedance gain, $k = 0.05$ Ω/A, (b) virtual impedance gain, $k = 0.15$ Ω/A

The following conclusion can be drawn from the above observations.

1. With the increase in virtual impedance gain, *k* the voltage stiffness of the inverter increased. As a result, the inverter injects lesser reactive power into the grid under a voltage sag.

2. The difference in reactive power outputs for two different virtual impedance gains is comparatively lesser at 5% voltage sag. The difference in reactive power outputs increases at higher voltage sag (i.e., 25%). The effect of the applied virtual impedance is more prominent at higher voltage sag than at lower voltage sag. It signifies that the voltage stiffness stays lower during lower voltage sags and increases at a much higher rate during higher voltage sags due to the effect of the proposed technique.

*C. Effect of virtual impedance on active power droop*

Finally, the effect of the proposed method on the active power droop characteristic which is responsible for frequency support, is examined by creating frequency dips from the grid emulator. The three-phase active power reference, $P^*$, is set to 1500 W. The performance of the inverter under two different frequency dips (from 50 Hz to 49.85 Hz and 49.7 Hz) with two different values of virtual impedance gain (0.05 Ω/A and 0.15 Ω/A) are observed. The active power outputs of the inverter in the mentioned conditions are shown in Fig. 15 and Fig. 16. The responses for two different virtual gains are nearly indistinguishable. The results indicate that the effect of the added virtual impedance is very little on the frequency droop characteristics. The proposed control technique successfully modifies the reactive power droop characteristics and hence the voltage stiffness without leaving any noticeable effect on the frequency support performance.

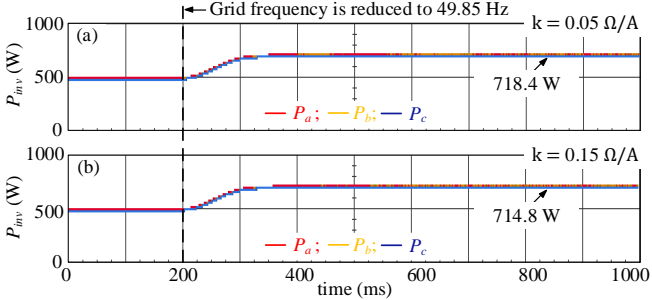

Fig. 15. Active power support by the inverter when the grid frequency reduces to 49.85 Hz from 50 Hz: (a) virtual impedance gain, $k = 0.05$ Ω/A, (b) virtual impedance gain, $k = 0.15$ Ω/A

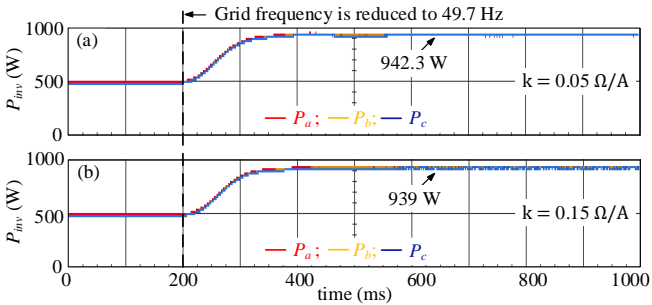

Fig. 16. Active power support by the inverter when the grid frequency reduces to 49.7 Hz from 50 Hz: (a) virtual impedance gain, $k = 0.05$ Ω/A, (b) virtual impedance gain, $k = 0.15$ Ω/A

## VI. CONCLUSION

This article primarily focused on achieving dynamic voltage stiffness control for a grid-forming inverter to improve its performance under weak grid situations. The concept of the Virtual Impedance (VIm) method is incorporated into a Virtual Oscillator (VO) based grid-forming control architecture to meet the mentioned objective. A detailed and systematic design procedure with analytical reasoning is presented. This article has also introduced a rigorous approach for the stability analysis of a VO-based grid-forming controller. The range of the control parameters can be obtained using the mentioned stability analysis. Along with the primary objective, the proposed concept of integrating the VIm method into a VOC has further applications, such as improving reactive power sharing among parallelly connected inverters. The proposed design and stability analysis approach is general and valid for other applications like the one mentioned above.

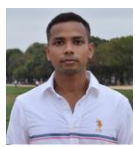
**Ritwik Ghosh** has completed M-Tech in Power Systems Engineering. Currently, he is pursuing his Ph.D. from Indian Institute of Technology Mandi. His research interest includes grid-forming control, non-linear oscillator based controller, and data-driven power system control.

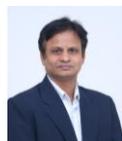
**Narsa Reddy Tummuru** (S'12, M'15, SM'20) received the B.Tech. degree in Electrical and Electronics Engineering from Jawaharlal Nehru Technological University, Hyderabad, India, in 2002, the M.Tech. degree in power electronics, electrical machines, and drives from the Indian Institute of Technology Delhi, New Delhi, India, in 2006, and the Ph.D. degree in electrical engineering from the Indian Institute of Technology Madras, Chennai, India, in 2015. From 2015 to 2016, he was a Postdoctoral Fellow with Nanyang Technological University, Singapore. He is currently an Associate Professor with the School of Computing and Electrical Engineering, Indian Institute of Technology Mandi, Kamand, India. His research interests include hybrid energy storage applications in smart grids, power electronic converter applications in renewable energy systems, power quality, control of switch-mode power converters, and wireless power transfer in electric vehicle applications. Dr. Tummuru received the Keshav-Rangnath Excellence and POSOCO Research Awards in the year 2015 and 2016, respectively, for his research work. He also received JSPS fellowship award in 2019. Dr. Reddy has been an Associate Editor of the IEEE ACCESS JOURNAL since 2019.

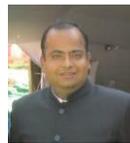
**Bhatar Singh Rajpurohit,** (Senior Member, IEEE) received M.Tech. degree in power apparatus and electric drives from IIT Roorkee, Roorkee, India, in 2005, and the Ph.D. degree in electrical engineering from IIT Kanpur, Kanpur, India, in 2010. He is currently working as Professor with the School of Computing and Electrical Engineering, IIT Mandi, Mandi, India. His major research interests include electric drives, renewable energy integration, and intelligent and energy-efficient buildings. He is a member of the International Society for Technology in Education, the Institution of Engineers (India), and the Institution of Electronics and Telecommunication Engineers.